\documentclass[12pt]{iopart}

\expandafter\let\csname equation*\endcsname\relax
\expandafter\let\csname endequation*\endcsname\relax

\usepackage{iopams}
\usepackage{cite}
\usepackage{graphicx}
\usepackage{mathtools}
\usepackage{physics}
\usepackage{comment}

\newcommand{\be}{\begin{equation}}
\newcommand{\ee}{\end{equation}}

\newcommand{\bpm}{\begin{pmatrix}}
\newcommand{\epm}{\end{pmatrix}}
\newcommand{\bea}{\begin{eqnarray}}
\newcommand{\eea}{\end{eqnarray}}
 
 \newcommand{\wD}{\widetilde{D}}
  \newcommand{\wE}{\widetilde{E}}
 
   \newcommand{\la} {\langle}
   \newcommand{\ra} {\rangle}
 \newcommand{\al}{\alpha}
\begin{document}
\title{Towards a Matrix Product Ansatz in Two Dimensions}
\author{Chandraniva  Guha Ray$^{1,2,3,4}$, 
        Aikya Banerjee$^{4,5}$, 
        P. K. Mohanty$^{1,4,\ast}$}
        
\address{$^1$Max Planck Institute for the Physics of Complex Systems, Nöthnitzer Straße 38, 01187 Dresden, Germany}
\address{$^2$Max Planck Institute of Molecular Cell Biology and Genetics, Pfotenhauerstraße 108, 01307 Dresden, Germany}
\address{$^3$Center for Systems Biology Dresden, Pfotenhauerstraße 108, 01307 Dresden, Germany}
\address{$^4$Department of Physical Sciences, Indian Institute of Science Education and Research Kolkata, Mohanpur, 741246 India}
\address{$^5$Department of Physics, University of Oxford, Denys Wilkinson Building, Keble Road, Oxford OX1 3RH, United Kingdom}
\ead{pkmohanty@iiserkol.ac.in}
\begin{abstract} 
Matrix product ansatz (MPA) is a powerful   framework  for constructing exact steady state  weights  of one dimensional non-equilibrium stochastic processes; but its generalization to higher dimensions is limited.       
Here, we introduce the MPA  formalism for two dimensions (2D). As a concrete application, we introduce and exactly solve a non-conserved assisted exclusion model (NAEM) in one and two dimensions with constrained hopping and local birth–death dynamics: a particle can hop to a neighbouring site only when exactly one of its neighbouring sites is vacant, while creation and annihilation occur exclusively at sites whose neighbours are all occupied. The MPA yields exact steady-state weights and provides a systematic method to compute observables such as density moments and particle currents. 
In the particle-conserving limit, the system undergoes an absorbing phase transition at the critical density $\rho_c=\tfrac12$ with order-parameter exponent $\beta=3$.    We further show that the steady state of the NAEM maps exactly onto the well-studied hard-square lattice gas with nearest-neighbour exclusion, thereby providing a nonequilibrium dynamical route to realizing equilibrium states of constrained lattice gases. Our work generalizes matrix-product methods beyond one dimension, establishing  a systematic approach to exact solutions of interacting stochastic systems in 2D.
\end{abstract}\maketitle

\section{Introduction}
Understanding stationary states of driven many-body systems remains one of the central open problems of non-equilibrium statistical mechanics. 
In equilibrium, the principle of detailed balance guarantees that steady states are described by the Gibbs--Boltzmann measure, providing a unified framework based on free energy minimization and ensemble theory \cite{Tolman38, Boltzmann64}. 
Out of equilibrium, however, sustained external driving generically breaks detailed balance, leading to non-equilibrium steady states (NESS) characterized by persistent probability currents, entropy production \cite{Privman97, Zia95, Bertini15_review,book3}. Systems in NESS exhibit a wide range of collective phenomena that have no direct equilibrium analogue.  Even in one dimension, these systems can display phase transitions induced by boundary effects, something forbidden in equilibrium under short-range interactions \cite{phase_tran_1, Evans1998, phase_tran_3,book3}. In addition to critical phenomena, non-equilibrium systems show a wide variety of emergent behaviors, including long-range correlations \cite{Basu10_2,Karimipour99}, non-trivial current fluctuations \cite{Lazarescu15}, and striking features such as current reversal under parameter tuning \cite{Chatterjee2018}. This complexity arises from the absence of a unifying principle akin to the free energy in equilibrium systems.
Even simple driven lattice gases can display boundary-induced phase transitions, long-range correlations despite short-range interactions, anomalous fluctuation statistics, and spontaneous formation of shocks or phase-separated structures \cite{Derrida2007, Basu10, Antal2000}. 
Such systems serve as minimal models for transport processes encountered in biology  \cite{Chowdhury2024}, traffic flow \cite{Hinsch} and granular flow \cite{Gaber2024}. 

Because no general framework comparable to equilibrium statistical mechanics exists, progress in the field has relied heavily on exact solutions of specific models. 
Several analytical approaches have proven successful, including coordinate and algebraic Bethe ansatz techniques \cite{bethe3,bethe2a,bethe2b}, transfer-matrix constructions \cite{Baxter1982}, and large-deviation methods that characterize fluctuations beyond average behavior \cite{Touchette2009, Bertini15_review}. 
Among these developments, one of the most influential advances has been the emergence of the \emph{Matrix Product Ansatz} (MPA).

The MPA first appeared in the exact solution of the open Totally Asymmetric Simple Exclusion Process (TASEP) \cite{open_TASEP1,open_TASEP2}. 
In this model, particles enter and leave a one-dimensional lattice while hopping subject to an exclusion constraint. 
Remarkably, the stationary probability of any configuration can be written as a product of non-commuting matrices associated with local occupation variables \cite{Derrida1993}. 
This representation transforms the problem of solving a stochastic dynamics into an algebraic problem defined by quadratic relations between matrices. Following this breakthrough, matrix product constructions were extended to a wide class of interacting particle systems \cite{gen_tasep_3,gen_tasep_4,gen_tasep_6,gen_tasep_6b,gen_tasep_7,gen_tasep_7b, Chatterjee2015,Aneva2016, Chatterjee2016a}, including multi-species exclusion processes \cite{Basu10}, systems with extended objects \cite{Gupta2011, Chatterjee2016}, assisted exclusion process  \cite{Basu2009, Chatterjee2017a},
particles without hardcore constraints \cite{Basu10_2, Chatterjee2017} and disordered systems \cite{Evans1997}. 
These exact solutions enabled explicit calculations of currents, density profiles, correlation functions, and fluctuation properties, revealing mechanisms behind symmetry breaking, shock localization, and phase coexistence in driven systems. 
Subsequent formal developments clarified the algebraic structure underlying MPA and established systematic criteria for its applicability \cite{Blythe07}. 

Despite its success, the Matrix Product Ansatz is intrinsically one-dimensional, where  correlations can be ordered sequentially; this  allows  the steady states to be expressed as products of matrices locally. 
In two dimensions, correlations spread along multiple directions, and no natural ordering preserves locality while maintaining algebraic closure. 
Consequently, extending MPA to higher-dimensional stochastic systems remains a major conceptual challenge. 
Although tensor-network generalizations such as projected entangled-pair states describe equilibrium and quantum systems \cite{Verstraete2008}, an exact analogue for non-equilibrium steady states of classical driven dynamics is still lacking. Establishing such a framework would open new analytical routes to transport, correlations, and phase behavior in two-dimensional driven systems.

In this work, we take steps toward a Matrix Product Ansatz in two dimensions. 
We formulate a framework aimed at capturing steady-state measures of two-dimensional classical stochastic models, analyze structural obstacles that prevent straightforward generalization from one dimension, and identify algebraic and geometric principles that may guide future constructions of higher-dimensional matrix product representations.

\section{Matrix Product Ansatz in One Dimension}

The Matrix Product Ansatz (MPA) has become a well-developed and powerful framework for obtaining exact steady states of one-dimensional driven stochastic systems. 
Comprehensive reviews and systematic formulations are available in the literature \cite{Blythe07}. 
The central idea of the MPA is to represent the stationary probability of a many-body configuration as a product of non-commuting matrices associated with local site variables, thereby transforming the steady-state condition into a set of algebraic relations between these matrices. 
Although the formalism is by now standard, we briefly review it here using a concrete example. 
This discussion establishes notation and highlights structural features that will be essential for formulating a Matrix Product Ansatz in two dimensions in the following section.

We consider a one-dimensional periodic lattice consisting of $L$ sites labeled by $i=1,2,\dots,L$. 
Each site carries a binary occupation variable $n_i\in\{0,1\}$, where $n_i=1$ denotes an occupied site and $n_i=0$ denotes a vacancy. 
The dynamics obeys a hard-core exclusion constraint, allowing at most one particle per site. 
To illustrate the construction of the Matrix Product Ansatz, we introduce a specific three-site interaction dynamics,
\be 
110  ~\xrightarrow[]{r} 101; ~~011  ~\xrightarrow[]{r} 101;
101~~\xrightleftharpoons[q]{p} ~~111
\label{eq:1D_NAEM}
\ee

A particle can hop from a given site to a neighbouring site only if the target site is the \underline{only} neighbouring site which is empty, i.e., the other neighbouring site must be  occupied. 
Thus, particle motion is \textit{assisted}, in the sense that hopping requires the presence of a neighbouring particle. In addition, a particle may be created or annihilated at a given site  with rates $p$ and $q$ respectively, only  
when both its  neighbouring sites are occupied by a particle.  Clearly   the  model  $p=0=q$   reduces to the   well-known  assisted  exclusion models \cite{deOliveira2005, Basu2009, Gabel2010}.  The birth death dynamics  violate density  conservation and we refer to this  model as non-conserved assisted exclusion model (NAEM). Our aim in this section would be to  introduce   and calculate  explicitly the  steady state weights  $W(\{ n_i\}$  of  configuratioins $\{ n_i\}$ of the   non-equilibrium dynamics given  in Eq.  \ref{eq:1D_NAEM} using Matrix Product Ansatz.

In the Matrix Product Ansatz (MPA), the steady-state weight of a configuration 
$\{n_i\}$ is written as
\begin{equation}
P(\{n_i\})=\Tr\!\left(\prod_{i=1}^{L} X_{n_i}\right),
\end{equation}
where a matrix $X_{n_i}$ is associated with the local state of site $i$.

If the steady state were a product measure, no spatial correlations would exist between site variables. 
Generic non-equilibrium steady states, however, exhibit non-trivial correlations and therefore cannot be expressed as simple product measures. 
The matrix product representation incorporates these correlations through the non-commutativity of the matrices $X_{n_i}$, while retaining many computational advantages of product states.
As a consequence, global quantities can be computed straightforwardly. 
For instance, the normalization (partition function) reads
\begin{equation}
\mathcal Z=\sum_{\{n_i\}} P(\{n_i\})
= \Tr\!\left[\left(\sum_{n} X_n\right)^L\right].
\end{equation}

For binary occupation variables $n_i\in\{0,1\}$, two matrices $X_0$ and $X_1$ represent vacant and occupied sites, respectively. 
For example, the configuration $\{\dots 10100110\dots\}$ has steady-state weight
$
\Tr\left(\cdots X_1 X_0 X_1 X_0 X_0 X_1 X_1 X_0 \cdots \right).
$
The partition function then simplifies to
$
\mathcal Z=\Tr\!\left[(X_0+X_1)^L\right].$ 
For notational convenience, we denote the matrices associated with occupied and vacant sites by $X_1 \equiv D$,  and $ X_0 \equiv E$. The probability vector $|P(t)\rangle$, evolving according to the master equation
\begin{equation}
\frac{d}{dt}|P(t)\rangle = M |P(t)\rangle ,
\end{equation}
can then be represented in matrix product form
\begin{equation}
|P\rangle =
\begin{pmatrix}
\Tr(DDD\cdots DD)\\
\Tr(DDD\cdots DE)\\
\Tr(DDD\cdots ED)\\
\Tr(DDD\cdots EE)\\
\vdots\\
\Tr(EEE\cdots EE)
\end{pmatrix},
\end{equation}
where each component corresponds to the steady-state weight of a configuration generated by replacing local states with matrices $D$ and $E$.

In the steady state, the condition $M|P\rangle = 0$
must be satisfied. Using the matrix product representation, this requirement can be written schematically as
\begin{equation}
M\,\Tr\!\left[A \otimes A \otimes \cdots \otimes A\right]=0,
\label{eq:ss_condition}
\end{equation}
where, for compact notation, we define
$A \equiv (D,E).$
Equation~\eqref{eq:ss_condition} encodes the stochastic dynamics through the transition rates and translates the steady-state master equation into algebraic relations that the matrices $D$ and $E$ must satisfy for a given dynamics.

The steady-state condition imposed by the master equation leads to a highly nontrivial constraint. 
In principle, one must satisfy $2^L$ coupled conditions, making it far from obvious how to determine a simple algebra for the matrices $D$ and $E$. 
The key simplification arises from a cancellation mechanism based on the locality of the dynamics.

For stochastic dynamics involving local interactions over $k$ consecutive sites, the Markov generator can be written as a sum of local operators,
\begin{equation}
M=\sum_{i} M_i ,
\end{equation}
where each term acts nontrivially only on sites $i,i+1,\dots,i+k-1$. 
Explicitly,
\begin{equation}
M_i=\mathcal{I}\otimes\mathcal{I}\otimes\cdots
\otimes \mathcal{K}
\otimes\cdots\otimes\mathcal{I},
\end{equation}
with $\mathcal{K}$ the local transition matrix describing the $k$-site dynamics, while the identity operator $\mathcal{I}$ acts on all remaining sites, leaving them unaffected.

 Now Eq. (\ref{eq:ss_condition}) reduces to a  simple condition
\be 
 \Tr[\mathcal{K} (A \otimes A \otimes \dots  \otimes A)_{k-{\rm terms}}] =0.
\ee
 This can be  achieved  using a telescopic (pairwise) cancellation. Say for a local $4$-site    dynamics,  if  we introduce  auxiliary matrices   $\widetilde{A} \equiv \begin{pmatrix}
   \widetilde{D} \\\widetilde{E}  
\end{pmatrix}$  so that 
\begin{equation}
\label{eq:cancel4}
\mathcal{K} (A \otimes A \otimes A \otimes A) = (  \widetilde{A} \otimes \widetilde{A} \otimes \widetilde{A}\otimes A - A \otimes  \widetilde{A} \otimes \widetilde{A}\otimes \widetilde{A}),
\end{equation}
then Eq. (\ref{eq:ss_condition})  will hold   automatically  for any choice of  $\widetilde{A}.$
The action of the local operator $M_i$ generates boundary terms involving sites near $i$, which appear with opposite signs when the neighbouring operator $M_{i+1}$ acts. 
Consequently, contributions produced at site $i$ are canceled by those generated at site $i+1$. 
Upon summing over all lattice positions, these terms cancel pairwise, yielding a telescopic sum and ensuring that the global steady-state condition $M|P\rangle=0$ is satisfied.

For the dynamics defined in Eq.~\eqref{eq:1D_NAEM}, we argue that the steady state contains no consecutive vacancies. 
Consequently, the effective steady-state dynamics reduces to a four-site process,
\begin{equation}
1101 \xrightleftharpoons[r]{r} 1011, 
\qquad
1010 \xrightleftharpoons[q]{p} 1110,
\qquad
1011 \xrightleftharpoons[q]{p} 1111 .
\label{eq:1D_NAEM_ss}
\end{equation}

\noindent \textit{Proof.}
From Eq.~\eqref{eq:1D_NAEM}, the particle-conserving part of the dynamics can be written as
\begin{equation}
110n \xrightarrow[]{r} 101n, 
\qquad 
n011 \xrightarrow[]{r} n101,
\qquad n=0,1 .
\end{equation}
In particular, configurations containing consecutive vacancies evolve according to
\(
1100 \xrightarrow[]{r} 1010 ,
\)
which eliminates a pair of adjacent zeros. Hence the total number of consecutive vacancies,
denoted by $N_{00}$, can only decrease under the dynamics. It is straightforward to verify that the non-conserving part of the dynamics does not create new $00$ pairs. 
Therefore, (a) in steady state, one must have $N_{00}=0$, (b) the particle density satisfies $\rho=\langle N\rangle/L \ge \tfrac{1}{2}$, and (c)
the effective steady-state dynamics reduces to Eq.~\eqref{eq:1D_NAEM_ss}, which contains no configurations with consecutive vacancies.

For the four-site dynamics given in Eq.~\eqref{eq:1D_NAEM_ss}, 
we construct the local $2^4\times 2^4$ transition matrix $\mathcal{K}$ 
and apply it to the cancellation condition Eq.~\eqref{eq:cancel4}. 
This procedure yields the following algebraic relations:
\bea
r( D^2ED- DED^2) =\wD\wE\wD D - D \wE\wD^2 = -(\wD^2 \wE D - D\wD \wE\wD)\cr 
pD^2ED- qD^4 = \wD^3 D - D\wD^3 = -(\wD^2 \wE D-D \wD\wE\wD)\cr
p EDED- q ED^3 = \wE\wD^2D - E \wD^3= - (\wE\wD\wE D - E \wD \wE\wD).
\eea
The remaining task is to determine whether matrices $D,E$ and auxiliary operators
$\widetilde D,\widetilde E$ exist that satisfy these relations.
Assuming the auxiliary operators reduce to scalars,
 $\wD=d, \wE=e$ consistency of the above equations requires $d=-e$.
The algebra then simplifies to
\bea
D^2ED-DED^2=0=p D^2ED-q D^4,\cr
p EDED- qED^3 =e^2(D+E).  \label{eq:E0}
\eea
At this stage, we are free to choose $e$, and there is no loss of generality in setting $e = 0$. 
Furthermore, in order to ensure that $N_{00} = 0$ in the steady state, an additional constraint must be imposed, namely $E^2=0.$ The resulting matrix algebra finally takes the form
\be 
E^2=0;~~ D^2 = \alpha D; ~~  (ED)^2  =  \alpha ED ~~ {\rm with}~\alpha=\frac{p}{q}.
\label{eq:algebra1D}
\ee

We now proceed   to find  a  representation. A simple two  dimensional representation  that obey  
the matrix algebra \eqref{eq:algebra1D} is  
 
\begin{align}
  D = \begin{pmatrix}
        \alpha & \alpha \\
        0 & 0 
        \end{pmatrix}= \alpha (\ket{1}\bra{1}+\ket{1}\bra{2}); \quad 
 E =  \begin{pmatrix}
        0 & 0 \\
        1 & 0 
        \end{pmatrix} = \ket{2}\bra{1}.
        \label{eq:representation}
\end{align}
These matrices provide a   exact steady state  solution  of  dynamics  Eq. \eqref{eq:1D_NAEM}.
The partition function of NAEM  in 1D  is then  $Z_{1D} = \Tr( T^L)= \lambda_+^L +\lambda_+^L$  where $T=D+E$ and  
$\lambda_\pm = \dfrac12\left(\alpha \pm  \sqrt{\alpha (4+\alpha)}\right)$ are  eigenvalues of $T.$ 
In the thermodynamic limit, the partition function  is  $Z_{1D} = \lambda_+^L$  and the  average   particle density of the system in steady state  is
\be 
\rho = \lim\limits_{L \to \infty}\frac{\Tr(DT^{L-1})}{\Tr(T^L)} =\frac12 \left(1+ \sqrt{\frac{\alpha} {4+\alpha}}\right). 
\ee
Naturally, the density of the system $\rho$ increases with  increase of  particle deposition rate $\alpha.$ The minimum density, obtained for $\alpha\to0$ is $\rho_c=1/2.$ This is consistent with the  fact that  the density of the system  cannot be less than $1/2$ as expected.

\section{Matrix Product  ansatz in two-dimension}
\begin{figure*}[t]
    \centering
    \includegraphics[width=15.8cm]{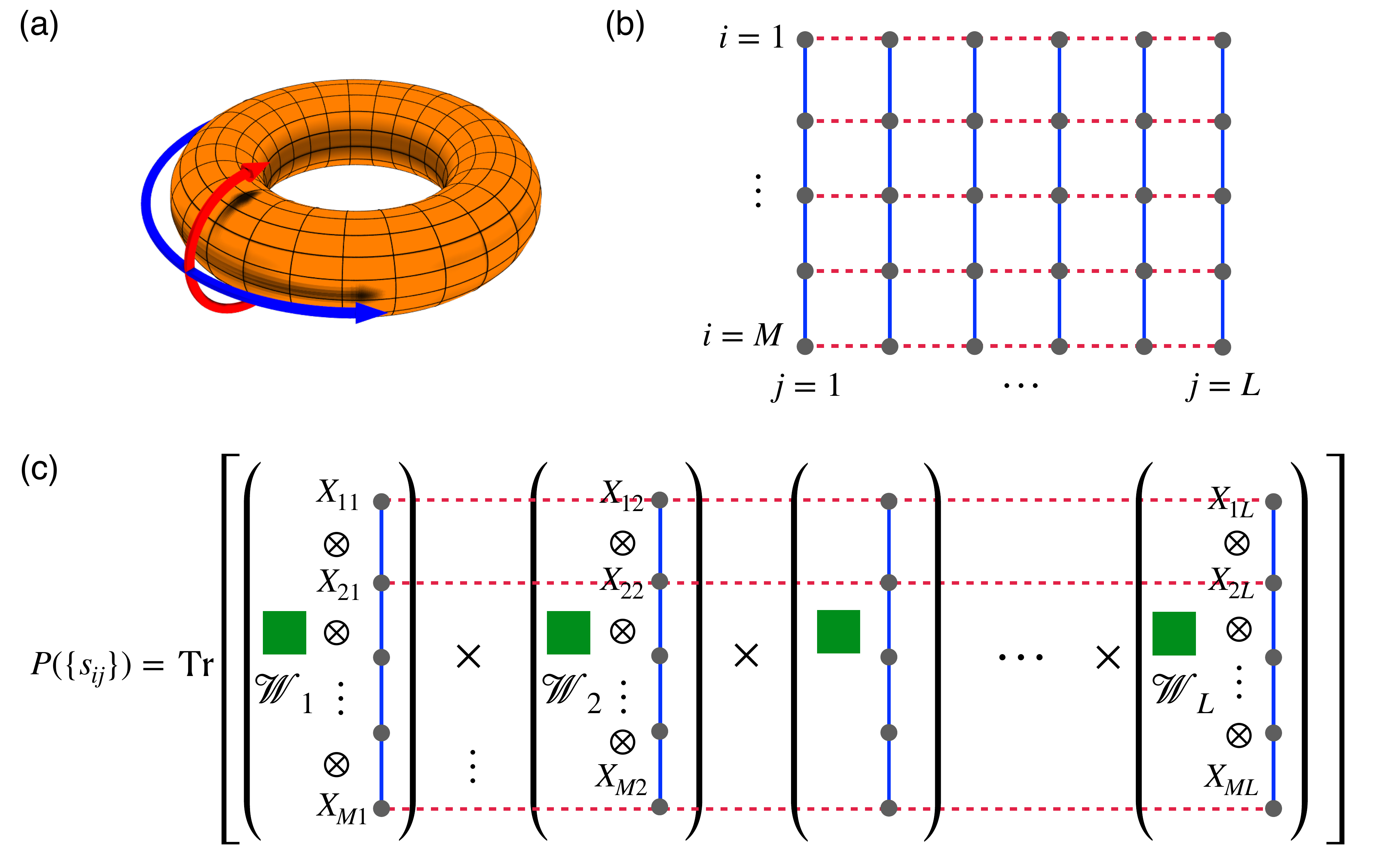}
    \caption{Schematic representation of MPA in 2D.
    A $M\times L$ square  lattice   with periodic boundary condition, like (a),  is composed of $L$ rods  of $M$-sites each,  connected by  interactions  (dashed lines in (b)).  Each of  the $j$-th rod  $\{ n_{1j}, n_{2j}, \dots, n_{Mj}\}$  is represented by a string of  matrix direct product $X_{1j}{\otimes} X_{2j}{\otimes}\dots  X_{Mj}$   of weight  $\mathcal{W_j}$ where  $X_{ij}$ is  a short-hand  for  $X_{s_{ij}}.$  The matrix direct product string $\prod^\otimes_i X_{ij}$  is   written  vertically, for visual appeal. The weight $\mathcal{W_j}$ depends on the configuration  of $j$-th rod, can be written, for an isotropic dynamics as  ${\mathcal W}_j = \Tr[X_{1j} X_{2j}\dots  X_{Mj}].$     }
    \label{figure1}
\end{figure*}
In this section, we introduce the two-dimensional matrix-product ansatz. In one dimension, the weight of a configuration is expressed as the trace  of a matrix string obtained by replacing each site variable with a  corresponding matrix; the resulting weight follows naturally from ordinary  matrix multiplication. Extending this construction to two dimensions is  nontrivial, since assigning matrices independently to each lattice site 
does not lead to a well-defined matrix operation.

To overcome this difficulty, we consider an $M\times L$ periodic square  lattice composed of $L$ vertical rods, each containing $M$ site variables and coupled through inter-rod interactions 
[see Fig.~\ref{figure1}(a),(b)]. The $j$-th rod, specified by the set of site variables $\{n_{1j},n_{2j},\dots,n_{Mj}\}$, is represented by the tensor-product string
$X_{1j}\otimes X_{2j}\otimes \cdots \otimes X_{Mj},$
so that each rod corresponds to a $2^{M}\times 2^{M}$ matrix (for binary variables $n_{ij}=0,1$). We further associate with each rod a configuration-dependent weight
${\cal W}_j \equiv 
{\mathcal W}(\{n_{1j},n_{2j},\dots,n_{Mj}\}).$
The matrix-product ansatz for the two-dimensional system is then defined as follows: the steady-state weight $P(\{n_{ij}\})$ of a generic configuration $\{n_{ij}\}$ can be written as
\be
P\left( \begin{matrix}
    n_{11} & n_{12}  &\ldots& n_{1L} \\
    n_{21} & n_{22} &\ldots& n_{2L} \\
    \vdots &\vdots &\ddots&\vdots \\
    n_{M1} & n_{M2} &\ldots& n_{ML}
\end{matrix}\right)
= \Tr\left[
 \begin{matrix}
    &X_{11} &&X_{12}  &\ldots&& X_{1L} \\
    &\otimes &&\otimes  &\ldots&& \otimes \\
    &X_{21} && X_{22} &\ldots&&X_{2L} \\
   {\cal W}_1& \otimes &{\cal W}_2& \otimes  &\ldots&{\cal W}_L& \otimes \\  
   &\vdots &&\vdots &\ddots&&\vdots \\
    &X_{M1} && X_{M2} &\ldots&& X_{ML}
\end{matrix}\right]
\label{eq:the_ansatz}
\ee
where   the weight of the rod for a system  with isotropic  dynamics  is   given by 
\be 
{\mathcal W}(\{ n_{1j}, n_{2j}, \dots, n_{Mj}\}) = \Tr( X_{1j} X_{2j} \dots X_{Mj} )\equiv {\mathcal W}_j.
\ee
A schematic   representation of the ansatz is   given in  Fig.  \ref{figure1}(c).  For example,  the steady state   weight  of  a  specific  configuration   of $3\times4$ system  is 
\be 
P\left(\begin{matrix}
    1 & 1 & 0 &1 \\
    0 & 1 & 0 &0 \\
    1 & 1 & 1 &1
\end{matrix}\right)
= \Tr\left[
\begin{matrix}
    &D && D && E &&D \\
    &\otimes&& \otimes&& \otimes&& \otimes\\
  \Tr(DED)  &E &\Tr(D^3)& D &\Tr(E^2D)& E &\Tr(DE^2)&E \\  
    &\otimes&& \otimes&& \otimes&& \otimes\\
    &D && D && D &&E
\end{matrix}\right] \nonumber
\ee
\be
=  \Tr(DED)\Tr(D^3)\Tr(E^2D)\Tr(DE^2) \Tr(D^ED)\Tr(EDE^2)\Tr(D^3E).
\ee
In deriving the last  step, we use the fact that  ${\rm Tr}(A\otimes B) ={\rm Tr}(A){\rm Tr}(B).$ Thus, the steady-state probability is just the product of traces of the corresponding matrices over all possible rows and columns, i.e., rods and chains. 

Here, we  describe the ansatz   for  a  isotropic dynamics  of a binary variable $n_{ij}=0,1$ on a square lattice. Extending  it to $n_{ij}=0,1, \dots, k$ or to other lattices is straight forward.   However, if the dynamics is not isotropic in $x$ and $y$ directions  we need two different set of matrices   to describe the rods (extending over $x$-direction)  and chains (extending over $y$-directions):    
\begin{align} \label{eq:2}
P(\{n_{ij}\}) = \prod_{j=1}^{L} {\rm Tr} \left(X_{{1j}}^{x}X_{{2j}}^{x}  \dots X_{{Mj}}^{x}\right) \prod_{i=1}^{M} {\rm Tr} \left( X_{{i1}}^{y}X_{{i2}}^{y} \dots X_{{iL}}^{y} \right).
\end{align}
 Then, the partition function for the steady state would be 
\begin{align} \label{eq:3}
    \mathcal{Z} = \sum_{\{n_{ij}\}} \left[\prod_{j=1}^{L} {\rm Tr} \left(\prod_{i=1}^{M} X_{{ij}}^{x} \right) \prod_{i=1}^{M} {\rm Tr} \left(\prod_{j=1}^{L} X_{{ij}}^{y} \right) \right].
\end{align}

We close this section with a notational remark. For visual clarity, the tensor-product string of matrices $\prod_i^{\otimes} X_{ij} \equiv X_{1j}\otimes X_{2j}\otimes \dots$ will occasionally be written in a vertical form, as written in Eq. \eqref{eq:the_ansatz}.

\section{Example model: non-conserved assisted exclusion}
\begin{figure}
    \centering
    \includegraphics[width=0.7\linewidth]{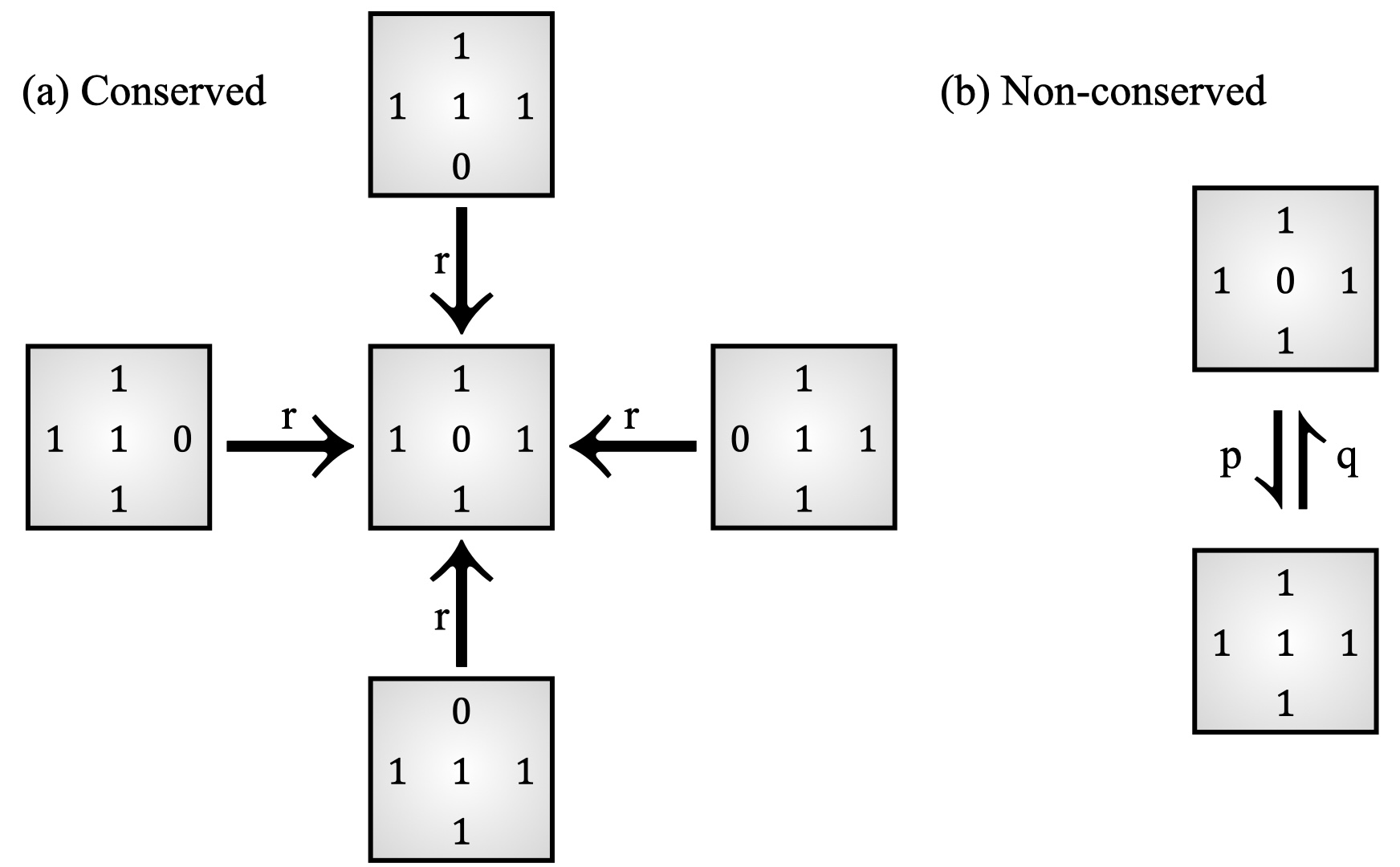}
    \caption{Dynamics of the non-conserved assisted exclusion model (NAEM) in 2D. (a) Conserved: a particle with exactly  one  neighbouring site  vacant,  move  there    at rate $r.$ (b) Non-conserved: a particle may be created or annihilated at rates $p$ and $q$ respectively,  only at sites  which have all its four neighbours  occupied.}
    \label{fig:dynamics1}
\end{figure}

\begin{figure}
    \centering
    \includegraphics[width=0.75\linewidth]{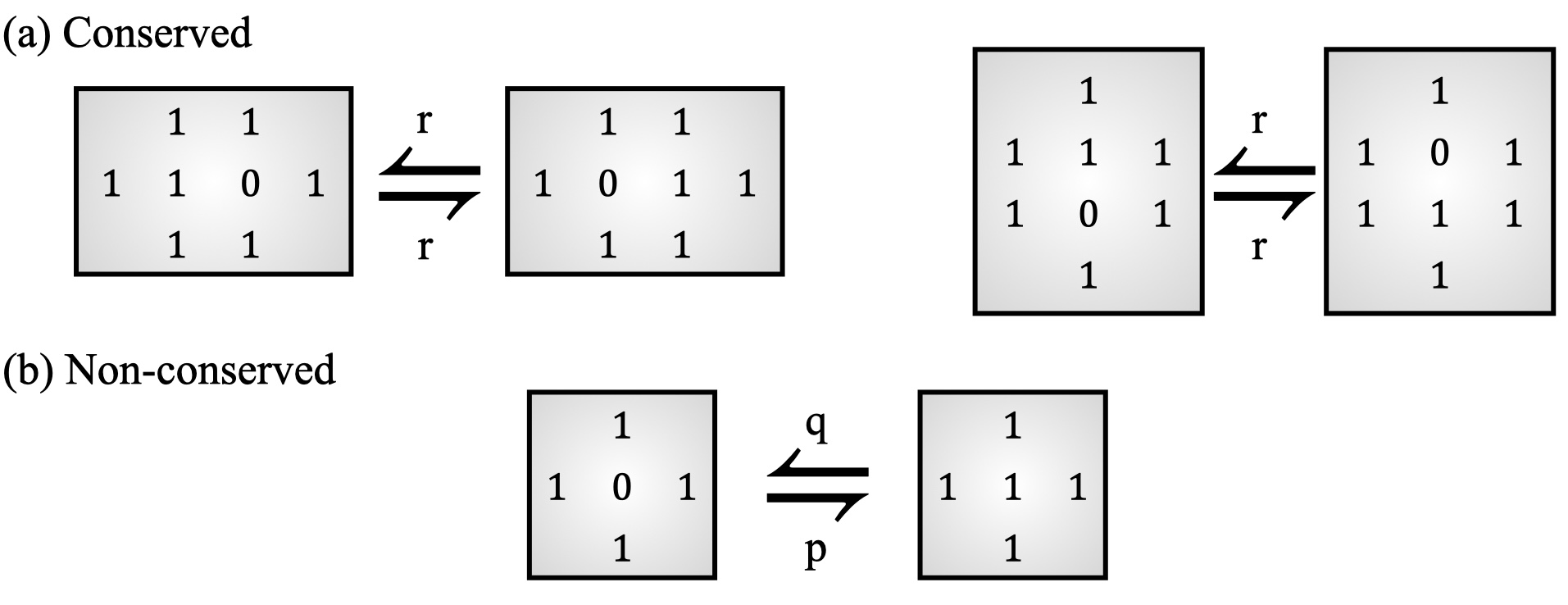}
    \caption{Steady-state dynamics of NAEM in 2D. In the steady state $N_{00}=0,$  and  thus,  neighbours of all vacant sites   are   occupied. The conserved part of the dynamics reduces  to  the following: Any $(10)$ pair  (in $x$ and $y$  direction) assisted by occupied neighbours can switch at rate $r.$}
    \label{fig:dynamics2}
\end{figure}

In order to illustrate a working example of a matrix product in 2D, we  extend the NAEM model to  two dimensions. In a $M\times L$ square lattice we assign integer variables $n_{ij}=0,1$ to each site ${\bf x}= (i,j)$  with $i=1,2, \dots,M$ and $j= 1,2,\dots,L.$. Regarding hard-core repulsion among particles, the sites can be vacant, $n_{ij}=0$, or occupied by at most one particle, ($n_{ij}=1$). These  hard-core particles  are  subjected to the following dynamics:  (a) a conserved  part in which   particles  having exactly  three occupied neighbours  move with  rate  $r$ to the only neighbouring site  which is vacant,  and (b) a non-conserved  dynamics  in which a  site  having all four neighbours  occupied   can  alter its  occupancy $n_{ij} \to 1-n_{ij}$   with rate $p+(q-p)n_{ij},$ i.e., a particle  is  added to a vacant site  with rate  $p$ and   occupied sites  are  emptied with rate $q.$   The dynamics of this  non-conserved  assisted exclusion model  (NAEM)  is  described   below schematically in Fig~\ref{fig:dynamics1}.
At first glance, the non-conserved dynamics might suggest that the model is ergodic, with all $2^{L^2}$ configurations accessible from any arbitrary initial state. However, this is not the case; like NAEM in 1D, here  too,  $N_{00}$ can  only decrease with time resulting in a steady state   with  $N_{00}=0$  which is equivalent to $\rho\ge \frac12$ (the proof is similar to 1D case). Thus, the steady state dynamics is  restricted only within a restricted configuration space ${\cal S},$ consisting of all configurations  with $N_{00}=0.$ Naturally the density of the system $\rho = \frac{\la N\ra}{L^2} \ge\frac12$   and  the steady state dynamics  is illustrated in Fig~\ref{fig:dynamics2}.

The next question is whether $\mathcal{S}$ is ergodic; that is, starting from any configuration $C \in \mathcal{S}$, can the dynamics reach any other configuration $C' \in \mathcal{S}$? The answer is yes, and a proof follows. Since $N_{00} = 0$ (meaning that every vacant site is surrounded by occupied neighbours), particles can be added to all vacant sites in any configuration $C \in \mathcal{S}$ to obtain a fully occupied configuration $C''$ with particle density $\rho = 1$. From this fully occupied state, any desired configuration $C' \in \mathcal{S}$ can then be constructed by removing particles from appropriate sites (which are occupied and have four occupied neighbours).

\subsection{A  matrix product steady state}
Since  the  dynamics of  NAEM is isotropic in $x$ and $y$ direction,    while  writing a   two dimensional   matrix product steady state, as described  in \eqref{eq:the_ansatz},   we set $X_1\equiv D$, $X_0\equiv E$   and make an ansatz  that the steady state  weight  of a configuration $C = \{ n_{ij}\}$ is 
\begin{align} \label{eq: W}
P(\{n_{ij}\}) = \prod_{j=1}^{L} {\rm Tr} \left(X_{{1j}}X_{{2j}}  \dots X_{{Mj}}\right) \prod_{i=1}^{M} {\rm Tr} \left( X_{{i1}}X_{{i2}} \dots X_{{iL}} \right). 
\end{align}
Let us remind   that   matrices  $X_{ij} \equiv   X_{n_{ij}}$  stands for the  $n_{ij} =0,1$ at site ${\bf x}=(i,j).$

The matrix-product weights must satisfy the steady-state condition
imposed by the Master equation corresponding to this specific dynamics.
A closer inspection of the conserved dynamics in Fig~\ref{fig:dynamics2} shows that the local update rule along the
direction of particle motion is
$1101 \;\xrightleftharpoons[r]{r}\; 1011 .$
For the non-conserved dynamics, acting in both the $x$- and
$y$-directions, the updates occur as $101 \;\xrightleftharpoons[q]{p}\; 111 .$
Taken together, the conserved and non-conserved processes in both
spatial directions are identical to the steady-state dynamical rules of
the one-dimensional NAEM. Therefore, it is sufficient to demonstrate
that the matrices $D$ and $E$ obey the same matrix algebra given in
Eq.~\eqref{eq:algebra1D}. Consequently, they admit the same matrix representation as obtained in Eq.~\eqref{eq:representation}: 
$D=\alpha (\ket{1}\bra{1}+\ket{1}\bra{2}), E=\ket{2}\bra{1}$  with $\alpha= p/q.$

To proceed further, we now need to determine the partition function of
the system. However, unlike in one dimension, an additional complication
arises here. Each occupation variable $n_{ij}$, and the corresponding matrix
$X_{ij}$, appears twice in the matrix-product construction: once as part
of the matrix weight of a rod (formally inside the trace) and once again
through the direct product structure defined along the rod itself.
This duplication originates from the geometry of the square lattice,
where every lattice \emph{site} is uniquely associated with the
intersection of two lines (or rods) passing through it. 
For the present problem, the statistical weight of a rod
$(n_{1j}, n_{2j}, \dots , n_{Mj})$ is  unity when there are no consecutive $0$s along the rod, and vanishes otherwise.  Hence the weight of the $j$-th rod can be written as
\be
{\cal W}_j
= \Tr( X_{1j} X_{2j} \dots X_{Mj})
= \prod_{i=1}^{M}\big[1-(1-n_{i,j})(1-n_{i+1,j})\big].
\ee
Our objective is therefore to construct a
$2^{M}\times 2^{M}$ transfer matrix $\mathcal{V}$ such that

\be
\mathcal{V} \begin{pmatrix}
               X_{1j}  \\            
               \otimes \\                
               X_{2j}\\ 
               \dots\\
               X_{Mj}          
\end{pmatrix} =\prod_{i=1}^M [1-(1-n_{i,j}) (1- n_{i+1,j})] \begin{pmatrix}
               X_{1j}  \\            
               \otimes \\                
               X_{2j}\\ 
               \dots\\
               X_{Mj}          
\end{pmatrix}
\label{eq:V}
\ee    
This  would help   us  obtain  the partition sum of the system  
explicitly, 
\bea
{\cal Z}  &=&  \sum\limits_{\{n_{ij}\}}  \prod\limits_{j=1}^{L}  {\rm Tr} \left(X_{{1j}}X_{{2j}}  \dots X_{{Mj}}\right) \begin{pmatrix}
               X_{1j}  \\            
               \otimes \\                
               X_{2j}\\ 
               \dots\\
               X_{Mj}          
\end{pmatrix}
=   \Tr[ \prod\limits_{j=1}^L\mathcal{V}\sum\limits_{n_{ij}} 
\begin{pmatrix}
               X_{1j}  \\            
               \otimes \\                
               X_{2j}\\ 
               \dots\\
               X_{Mj}          
\end{pmatrix}] \cr
&& =  \Tr[ \prod\limits_{j=1}^L\mathcal{V} 
\begin{pmatrix}
              \sum\limits_{n_{1j}} X_{1j}  \\            
               \otimes \\                
              \sum\limits_{n_{2j}} X_{2j}\\ 
               \dots\\
               \sum\limits_{n_{Mj}}X_{Mj}          
\end{pmatrix}]= \Tr[ \prod\limits_{j=1}^L\mathcal{V} 
\begin{pmatrix}
              T  \\            
               \otimes \\                
             T\\ 
               \dots\\
               T         
\end{pmatrix}]
=Tr[(\mathcal{V}\mathcal{T})^L]
\eea
where we use   $\sum\limits_{n_{ij}=0,1} X_{ij}=D+E= T$ and $\mathcal{T}= T^{\otimes M}.$ The success of this method   for  calculating the partition function   depends on Eq. \eqref{eq:V}, i.e., can we find a ${\mathcal V}$ which satisfies  Eq. \eqref{eq:V}? 
 Let us write 
\be 
{\cal V} = R_M \prod_{i=1}^{M-1} R_i,  ~~~~  {\rm with} ~~ R_{i}  = \bpm I_2\\\otimes \\I_2\\ \dots\\  R \\ \dots \\I_2\epm 
\begin{matrix} ~\\~ \\~\\ ~\\  \rightarrow {\rm at}~ i, {\rm covering}~i ~{\rm and}~ i+1 ~ {\rm position,}\\ ~~~~
\\~\end{matrix}\label{eq:Ri}
\ee
where  $R$ is a  $4\times 4$ matrix  so that  $R_i$  acts  on site $i$ and $i+1,$  keeping other  sites  unchanged. To account for  periodic boundary condition we have a separate   matrix $R_M,$ that acts on the $M$-th and  $1^{\rm st}$  site  to ensure that  both  of  them are not vacant. 

Clearly  for  Eq. \eqref{eq:V}  to hold , it is enough  to show that 
\be
R\bpm X_{ij}\\\otimes\\ X_{i+1,j}\epm  = \big[1-(1-n_{i,j})(1-n_{i+1,j})\big]\bpm X_{ij}\\\otimes\\ X_{i+1,j}\epm. 
\ee
Explicitly, 
\be
R\bpm D\\\otimes\\ D\epm  = \bpm D\\\otimes\\D\epm; ~~R\bpm D\\\otimes\\E\epm  = \bpm D\\\otimes\\E\epm; ~~ R\bpm E\\\otimes\\D\epm  = \bpm E\\\otimes\\D\epm; ~~R\bpm E\\\otimes\\E\epm  = 0.
\label{eq:R}
\ee
It is not difficult to verify that, indeed   such a matrix exists:
\be 
R=  \bpm  1&0&0&0\\0&1&0&0\\0&0&1&0\\0&0&0&0\epm =  \bpm I_2  &               & A\\
               \otimes & -             &  \otimes \\      
               I_2  &               & A  \epm, ~~~~  {\rm with}~~A =|2\rangle\langle2| = \bpm0&0\\0&1\epm.
\ee
The issue with \(R\) is that it cannot be written as a direct product of two \(2\times2\) matrices, which would have given a simpler form for \(R_M\). However, \(R_M\) can still be expressed as a sum of two matrices:
\be
R_M =I_2 ^{\otimes M} - A\otimes I_2^{\otimes (M-2)}\otimes A, 
\ee
which obey  
\be
R_M \begin{pmatrix}
               X_{1j}  \\            
               \otimes \\                
               X_{2j}\\ 
               \vdots\\
               X_{Mj}          
\end{pmatrix} = [1-(1-n_{1,j}) (1- n_{M,j})] \begin{pmatrix}
               X_{1j}  \\            
               \otimes \\                
               X_{2j}\\ 
               \vdots\\
               X_{Mj}          
\end{pmatrix}
\label{eq:R_M}
\ee  
To find a  more  convenient  form of ${\cal V}$ 
we notice that the set of matrices $\{R_i\}$ commute  with each other $[R_i,R_j] =0, ~ \forall i,j.$  Now we separate the products  for  even and odd  values  of $i$. For an even $M=2m$, we write 
\be
{\cal V} =\prod_{i=1}^M R_i=  {\cal V}_{o}  {\cal V}_{e}; ~~ {\cal V}_{o} =R^{\otimes m}; ~~{\cal V}_{e} =I_2\otimes R^{\otimes (m-1)}\otimes I_2 -A\otimes R^{\otimes (m-1)}\otimes A.
\ee

With this,   we  summarize the results   we obtained  using  MPA2D.    We have an exact partition  function  that describes the steady state  of  NAEM    on a  two dimensional  $2m \times L$ square lattice. The partition function   is, 
 \bea
{\mathcal Z} =\Tr \mathbb{T}^L;~~\mathbb{T}= {\mathcal V}_o {\mathcal V}_e \mathcal{T}; ~ {\rm with}~{\cal T}= T^{\otimes 2m};~~  {\cal V}_{o} =R^{\otimes m}; ~~{\cal V}_{e}={\cal V}_{e}^+ -   {\cal V}_{e}^-;\cr
{\rm  where}~ {\cal V}_{e}^+ =I_2\otimes R^{\otimes (m-1)}\otimes I_2;  {\cal V}_{e}^-=  A\otimes R^{\otimes (m-1)}\otimes A,
\eea
where the  associated matrices, with $\alpha=\frac{p}q,$ are  
 \be 
 T =\bpm \alpha &\alpha\\1&0\epm, ~~A =\bpm 0 &0\\0&1\epm ~~{\rm and }~~ R=  \bpm  1&0&0&0\\0&1&0&0\\0&0&1&0\\0&0&0&0\epm. 
\ee
The  expression of  the partition function is exact.  And in the thermodynamic limit, $L\to \infty$,  one  can write   ${\cal Z}  =  \Lambda_m^L$   where  $\Lambda_m=\boldsymbol{\rho} (\mathbb{T})$  is the  spectral radius (the largest among the absolute values of the eigenvalues of $\mathbb{T}$).
For small $m,$  one can calculate $\Lambda_m$ directly by brute-force diagonalization, which we will do in the  next  section. An approximate calculation of $\Lambda_m$ for large $M=2m$ is  done below. First we show 
(see the footnote\footnote{
The matrices ${\cal V}^\pm$ are diagonal with entries 
${\cal V}^\pm_{ii}\in\{0,1\}$. Moreover,
${\cal V}^-_{ii}=1 \;\Rightarrow\; {\cal V}^+_{ii}=1,$
since ${\cal V}^+$ is obtained from ${\cal V}^-$ by replacing 
$A=\ket{2}\!\bra{2}$ with the identity $I_2$.
If this were \emph{not} the case, i.e.\ if there existed an index $j$
such that ${\cal V}^-_{jj}=1$ but ${\cal V}^+_{jj}=0$, then
${\cal V}^-C$ would retain the $j$-th coordinate while
${\cal V}^+C$ would eliminate it. Consequently, if the dominant
eigenvector of $C$ were supported entirely on that coordinate,
one could have $\rho({\cal V}^+C)<\rho({\cal V}^-C).$
Fortunately, this situation does not arise here.
}) that 
\be 
 \boldsymbol{ \rho}({\mathbb T})=   \boldsymbol{ \rho}({\mathcal V}_e^+ {\mathcal V}_o {\mathcal T}- {\mathcal V}_e^- {\mathcal V}_o {\mathcal T}) \le    \boldsymbol{\rho}({\mathcal V}_e^+ {\mathcal V}_o {\mathcal T}).
\ee
Then for large $M=2m,$  one can  approximate $\Lambda_m \simeq \boldsymbol{\rho}({\mathcal V}_e^+ {\mathcal V}_o {\mathcal T}).$  The difficulty remain in multiplying  matrices  ${\mathcal V}_e^+,$  which is a $(m+1)$-factor  tensor product ($2\times4^{m-1}\times 2$),   with   ${\mathcal V}_o$  which is a $m$-factor tensor product ($4^m$), which can  not be simplified beyond a direct multiplication in $4^m$-space  unless matrix $R$  is  in a direct product form $R_1\otimes R_2$  with $R_{1,2}$  (which is not the case here). However, since we  want only the spectral radius, we  have 
\bea
\boldsymbol{\rho}({\mathcal V}_e^+ {\mathcal V}_o {\mathcal T}) = 
\boldsymbol{\rho}(\tilde {\mathcal V}_e^+ 
\tilde {\mathcal V}_o \tilde{\mathcal T}),~ {\rm with}~\tilde {\mathcal V}_o =  {\mathcal V}_o\otimes I_4; \tilde{\mathcal T}={\mathcal T}\otimes I_4,\cr
{\rm and}~~ \tilde {\mathcal V}_e^+ =  I_2 \otimes {\mathcal V}_e^+\otimes I_2= I_4 \otimes  R^{\otimes(m-1)}\otimes I_4
\eea
The task is now simplified. 
\be
\boldsymbol{\rho}({\mathcal V}_e^+ {\mathcal V}_o {\mathcal T}) =\boldsymbol{\rho}\left( R T_4 \otimes (R^2T_4)^{\otimes (m-1)} \otimes I_4\right)= \boldsymbol{\rho}(K)^{m};  
\ee
where  $T_4= T \otimes T$ and we use  the  fact $R^2 =R$ to write   
\be
K =R^2T_4 = RT_4= \bpm \alpha^2& \alpha^2& \alpha^2&\alpha^2\\
\alpha &0&\alpha &0\\ \alpha &\alpha &0&0 \\0&0&0&0
\epm.
\ee
The eigenvalues of $K$  are 
\be 
\left\{ 0, -\alpha, \lambda_\pm=\frac\alpha2\left( 1+\alpha  \pm \sqrt{1+ 6\alpha +\alpha^2}\right)\right\}
\ee
and  the spectral radius is  $ \boldsymbol{\rho}(K) =\lambda_+.$  Thus the  partition function is 
\be
{\cal Z} = {\rm Tr}( \mathbb{T}^L) \sim \boldsymbol{\rho}(K)^L = \lambda_+^{m L}=\left(\frac\alpha2\left[1+\alpha  + \sqrt{1+ 6\alpha +\alpha^2}\right]\right)^{\frac{ML}2} 
\label{eq:Z}
\ee

To calculate the steady-state values of the observables, we note from the structure of the matrix $D = \alpha \bpm 1 &1\\0&0\epm,$ that \(\alpha\) plays the role of a particle fugacity. If the canonical partition function of the steady-state configurations with exactly \(N\) particles is denoted by \(Q_N\), then \(\mathcal{Z}\) can be written as
\be 
{\cal Z} =   \sum\limits_{N=0}^\infty \alpha^N Q_N.
\label{eq:GCE}
\ee
Thus, by   writing $\alpha \equiv e^\mu$   we  obtain,  
\be \rho = \frac{\la N\ra}{M L} = \frac{1}{ML} \frac{2}{\cal Z} \frac{d {\cal Z}}{d\mu},  ~~{\rm and}~~ \rho_k= \frac{\la N^k\ra}{(ML)^k}
=\frac{1}{(ML)^k} \frac{2^k}{\cal Z}   \frac{d^k \cal Z}{d\mu^k}, ~~ k=2,3,\dots 
\label{eq:moments}
\ee
From Eq. \eqref{eq:Z},   the first three moments of the particle density are, 
\bea
\rho &=&(-1 + \al + 3 \eta)/4 \eta,\cr
\rho_2 &=&\left(-3 - 11 \al + 19 \al^2 + 3 \al^3 + \eta[5  + 26 \al  + 5 \al^2]\right) /8 \eta^2,\cr
\rho_3 &=&\left(-7 - 53 \al - 50 \al^2 + 302 \al^3 + 89 \al^4 + 7 \al^5  \right. \cr
 && ~~~~~~~~+\left. \eta[9  + 84 \al  + 234 \al^2  + 96 \al^3  + 9 \al^4 ]\right)/16 \eta^5
 \label{eq:rho23},
\eea
where $\eta = \sqrt{1+6\alpha + \alpha^2}$.
\begin{figure*}[t]
    \centering
    \includegraphics[width=16cm]{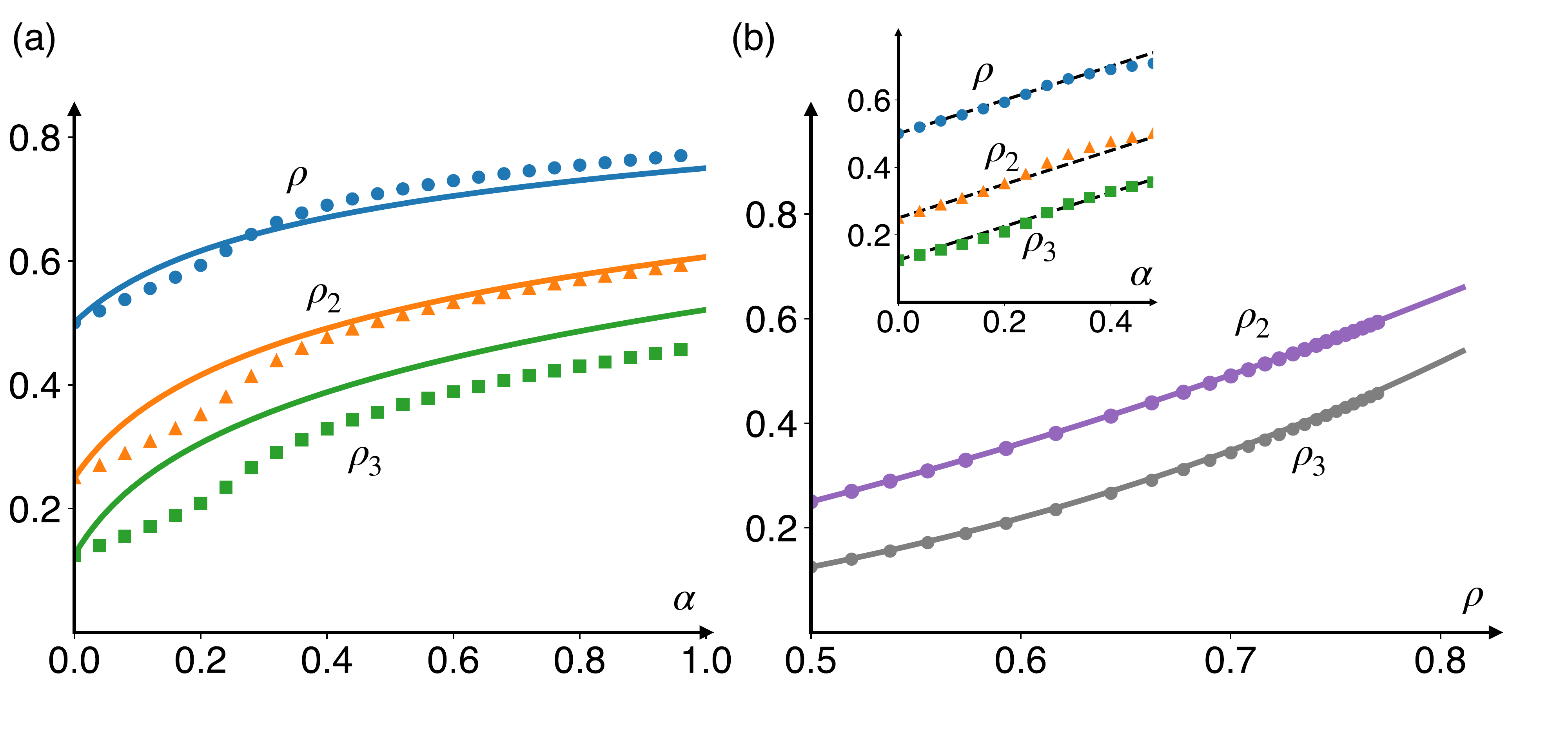}
    \caption{(a) $\rho,$  $\rho_2$ and $\rho_3$ as a function of $\alpha,$ computed from Monte Carlo simulations of NAEM in 2D  (symbols)  on  a  $20\times 100$ system  are compared with   Eq. \eqref{eq:rho23}.  (b) $\rho_2$ vs. $\rho$ and $\rho_3$ vs. $\rho$  obtained following Eq. \eqref{eq:Z2}  (lines) matched well with the  those obtained from  simulations.  Inset: for small $\alpha,$  $\rho,$  $\rho_2$ and $\rho_3$  are linear  in $\alpha$ as predicted  by Eq. \eqref{eq:linear}.}
    \label{fig:rho23}
\end{figure*}
In Fig. \ref{fig:rho23}(a), we plot the  first three  moments  obtained  from  simulations for a system of size $20 \times 100$ along with  the  approximate   density calculated in Eq.  \eqref{eq:rho23}, using MPA2D. The qualitative features of    $\rho, \rho_2, \rho_3$   are  found to be similar for  large $\al$; the approximation, however, fails badly  near $\alpha=0.$ For small $\al$,  Eq. \eqref{eq:rho23} give us  
\be
\rho = \frac12  + \al + {\cal O}(\al^2),~~ \rho_2 = \frac14  + \frac32 \al + {\cal O}(\al^2),~~\rho_3 = \frac18  +\frac74 \al + {\cal O}(\al^2),
\ee
whereas the simulations  show  that  the first order correction to  all three observables  $\rho, \rho_2, \rho_3$ near   $\al=0$  is  $\frac12 \al$ (dashed line in  \ref{fig:rho23}(a)).  We realize that this is due the fact that in linear order $\boldsymbol{\rho}(K)=\lambda_+ \sim \al,$  competes  with  the absolute value of the other eigenvalue $|-\al|.$    Thus the natural correction to the partition function  is, 
\be
{\cal Z} =  \boldsymbol{\rho}(K)^L = \left(\lambda_+^m + \alpha^m\right)^L
\label{eq:Z2}
\ee
Then some algebraic manipulations of Eq. \eqref{eq:moments}  give us  the moments  up to  the  leading order in $\alpha$ (in thermodynamic limit) to be  
\be
\rho = \frac12  + \frac12 \al + {\cal O}(\al^2),~~ \rho_2 = \frac14  + \frac12 \al + {\cal O}(\al^2),~~\rho_3 = \frac18  +\frac12 \al + {\cal O}(\al^2).\label{eq:linear}
\ee
The exact expressions for $\rho,\rho_k$ are rather lengthy and are therefore omitted here; however, they can be derived in a straightforward manner. In Fig. \ref{fig:rho23}(a), we have  plotted  $\rho_2$ and $\rho_3$ as functions of $\rho,$ along with  the results we  obtained from  ${\cal Z}$ in  Eq.~\eqref{eq:Z2}.  They match quite well.
  
\subsection{Exact results for small $M=2m$}
In this section, we explicitly calculate \(\mathcal{Z}\) for small systems with \(M=2m\).  
For \(m=1\), the system reduces to a \(2\times L\) ladder, and the dynamics given in Fig \ref{fig:dynamics1} cannot be implemented directly.  
The smallest nontrivial case is \(m=2\), where
\be 
{\cal Z} =\Tr \mathbb T^L,  ~~{\rm with}~~ {\mathbb T}=(I_2 \otimes R \otimes I_2 - A \otimes R \otimes A)^2 (R \otimes R  ~ T \otimes T)^2
\ee
being  a $2^4\times 2^4$ square matrix. The eigenvalues$\{\lambda\}$  of $\mathbb T$ obey  the characteristic equation
\bea
\lambda^9(\lambda+\al^3)^2 (\lambda^2+ (\al^2 + \al^3) \lambda -\al^5)(\lambda^3- 3\al_2 \lambda^2 -3\al_1\lambda +\al^5)=0,\cr 
{\rm where}~~ 
\al_2=\al^3 + (\al^2 +\al^4)/3,~~ {\rm and } ~~ \al_1=(\al^4 +\al^5-\al^3)/3.
\label{eq:charEq_T4}
\eea
Nine eigenvalues of \(\mathbb{T}_4\) vanish, reflecting the fact that, on the $M=4$ periodic lattice, 9 out of the 16 configurations contain at least one consecutive pair of zeros. The remaining 7 allowed configurations  are
\be
{\cal S}_4=\{0101, 0111, 1010, 1011,1101,1110,1111\}.
\label{eq:S4}
\ee which  is the   configuration space of   NAEM   on an 1D lattice   of size $M=4.$ This also implies that the transfer matrix \(\mathbb{T}\) is reducible and it admits a \(7\times7\) representation, denoted by \({\mathbb{T}}_4\), which acts on ${\cal S}_4.$ In  fact  for any generic $M$  we have a $f(M)$ dimensional  reduced matrix   where  $f(.)$  is a  Febonacci sequence 
 obeying  $f(n)= f(n-1) + f(n-2)$ with  $f(1)=1$ and $f(2)=3.$  Note that $f(2)=3$   corresponds to   the  three configurations    for $M=2$ : ${\cal S}_2= \{01,10,11\}$.

 For $M=4$   the transfer matrix is then 
\be
{\mathbb{T}}_4=\left(
\begin{array}{ccccccc}
 0 & 0 & \alpha ^2 & \alpha ^{2} & 0 & \alpha ^{2} & \alpha ^2 \\
 0 & 0 & \alpha ^{3} & \alpha ^3 & \alpha ^3 & \alpha ^3 & \alpha ^{3} \\
 \alpha ^2 & \alpha ^{2} & 0 & 0 & \alpha ^{2} & 0 & \alpha ^2 \\
 \alpha ^{3} & \alpha ^3 & 0 & 0 & \alpha ^3 & \alpha ^3 & \alpha ^{3} \\
 0 & \alpha ^3 & \alpha ^{3} & \alpha ^3 & 0 & \alpha ^3 & \alpha ^{3} \\
 \alpha ^{3} & \alpha ^3 & 0 & \alpha ^3 & \alpha ^3 & 0 & \alpha ^{3} \\
 \alpha ^4 & \alpha ^{4} & \alpha ^4 & \alpha ^{4} & \alpha ^{4} & \alpha ^{4} & \alpha ^4 \\
\end{array}
\right).
\label{eq:T4}
\ee 
One can easily check that  ${\mathbb{T}}_4$ has the same characteristic equation  given in Eq. \eqref{eq:charEq_T4}, except  the $\lambda^9$  term. The largest eigenvalue, $\Lambda$,   comes from the  solution of the cubic equation $\lambda^3- 3\al_2 \lambda^2 -3\al_1\lambda +\al^5=0$, 
\bea
\Lambda = \al_2 + v + \frac1{v}(\al_1+ \al_2^2),\cr
2v^3= u + \sqrt{u^2  -4(\alpha _2^2+\alpha _1)}, ~~ u=2 \alpha _2^3+3 \alpha _1 \alpha _2 -\alpha ^3.
\eea
The mean density of the system  is then, 
\be 
\rho= \frac{\al}{4L}  \frac{d }{d\al} \ln {\cal Z}=\frac{\al}{4\Lambda} \frac{d \Lambda}{d\al}.
\label{eq:densityexact}
\ee
\begin{figure*}[t]
    \centering
    \includegraphics[width=16cm]{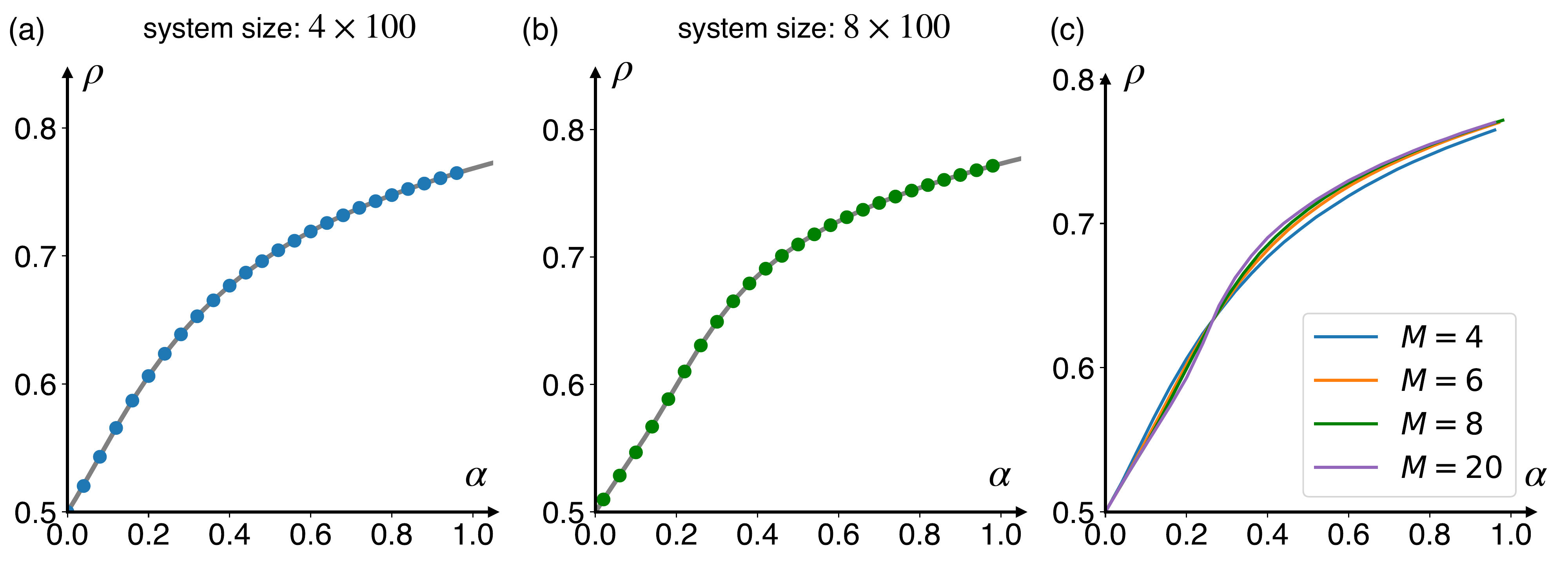}
    \caption{Density plots for $M = 2m$. Exact  results  ($\rho$ versus $\alpha$) for small $M$  following Eq. \eqref{eq:densityexact}  (solid lines), for  (a)  $4\times 100$ 
    and (b) $8\times 100$ are compared with  the  same obtained from   Monte-Carlo simulations (symbols).
    (c) Comparison of $\rho$ versus $\alpha$ for different system size, $M=4,6,8$  and $20$. $M=20$ is already in the large $M$ limit.}
    \label{fig:rho_M}
\end{figure*}
Again, we omit the exact but lengthy expression for the density at \(M=4\). In Fig.~\ref{fig:rho_M}(a), we compare the analytical result with the steady-state density obtained from Monte Carlo simulations of the model, showing very good agreement. Exact results for \(M=8\) are presented in Fig.~\ref{fig:rho_M}(b).
Note that for small \(M\), the functional form of \(\rho(\alpha)\) depends strongly on \(M\). A comparison of \(\rho(\alpha)\) for \(M=4,6,8,20\) with \(L=100\) is shown in Fig.~\ref{fig:rho_M}(c). As \(M\) increases, the curves approach a limiting behavior, which is already well approximated by the result for \(M=20\).

To calculate other observables, we employ a projector method described below.  
Consider the two-point correlation function $C_x(r)=\langle  n_{i,j} n_{i+r,j}\rangle,$
which represents the probability that both the \(i\)-th and the \((i+r)\)-th rods have a particle at the \(j\)-th site. Since \(C_x(r)\) is independent of \(i\) and \(j\), we can set \(j=M\) without loss of generality.
Let \(\mathbb{D}\) and \(\mathbb{E}\) denote matrices that project onto all configurations of \(\mathcal{S}_M\) having occupation \(1\) and \(0\), respectively, at the \(M\)-th site. For example, from Eq.~\eqref{eq:S4}, by inspection, we obtain for \(M=4\),
\bea
{\mathbb D}=|1\rangle\langle 1| + |2\rangle\langle 2| + |4\rangle\langle 4| + |5\rangle\langle 5| + |7\rangle\langle 7| \cr
{\mathbb E}=|3\rangle\langle 3| + |6\rangle\langle 6|= I_7 - \mathbb D.
\eea
Using  similar projectors  for generic $M,$ we get 
\bea
\rho = \frac{\Tr( {\mathbb DT}^L)} {\Tr( {\mathbb T}^L)}\simeq \frac{ \la w| \mathbb D|v\ra}{\la w|v\ra  }; ~~ \rho_0 = \frac{\Tr( {\mathbb ET}^L)} {\Tr( {\mathbb T}^L)}\simeq  \frac{\la w| \mathbb E|v\ra}{\la w|v\ra  };\cr
C_x(r)=\la n_{i,j} n_{i+r,j} \ra=  \frac{\la w| \mathbb {DT}^r \mathbb D|v\ra}{\Lambda^r\la w|v\ra  }, 
\label{eq:projector_method}
\eea
where \(\langle w|\) and \(|v\rangle\) are the left and right eigenvectors of \(\mathbb{T}\) corresponding to the largest eigenvalue \(\Lambda\), i.e., $\la w|\mathbb T = \la w|\Lambda$, $\mathbb T|v\ra = \Lambda|v\ra.$  Note that in the last step we have taken the limit \(L \to \infty\), where the largest eigenvalue \(\Lambda\) dominates over all other contributions.

An interesting observable is the current \(J_{x,y}\), which measures how often particles move in the positive $x$- (rightward) and positive \(y\)- (upward) directions, respectively. Since the dynamics obey detailed balance in the steady state, the magnitude of the current in the opposite directions (left and down) is also \(J_{x,y}\). Using the projector matrices, we can write
\be 
J_x=\Biggl\la  \begin{matrix} . &1&. \\1&1&0\\ . &1&.\end{matrix} \Biggr \ra = \frac{\Tr( \mathbb{DTD}_{111}\mathbb{TET} ^{L-2})} {\Tr( {\mathbb T}^L)}\simeq \frac{\la w| {\mathbb {DTD}_{111} \mathbb{TE}}|v\ra}{\Lambda^2\la w|v\ra  },
\ee
where  $\mathbb D_{111}$ is a projector that projects out rod-configurations    which  is  certainly occupied at    three places $j=1,M-1$ and $M.$  For $M=4,$ $\mathbb D_{111} =|4\rangle\langle 4| + |7\rangle\langle 7|.$  In a same way we obtain, 
\be
J_y=\Biggl\la  \begin{matrix} . &0&. \\1&1&1\\ . &1&.\end{matrix} \Biggr \ra =\frac{\Tr( \mathbb{DTD}_{011}\mathbb{TDT} ^{L-2})} {\Tr( {\mathbb T}^L)}\simeq \frac{\la w| {\mathbb {DTD}_{011} \mathbb{TD}}|v\ra}{\Lambda^2\la w|v\ra },
\ee
where $\mathbb D_{011}$ is a projector that ensures the first site of the rod is vacant, while the $(M-1)$-th and $M$-th sites are occupied. For $M=4$, $\mathbb D_{011}=|2\rangle\langle2|$. From these expressions, it is clear that $J_x$ and $J_y$ are not identical. This difference arises from the asymmetry between the $x$- and $y$-directions due to the rectangular geometry. As $M$ increases, this asymmetry gradually disappears.

\begin{figure*}[t]
    \centering
     \includegraphics[width=16cm]{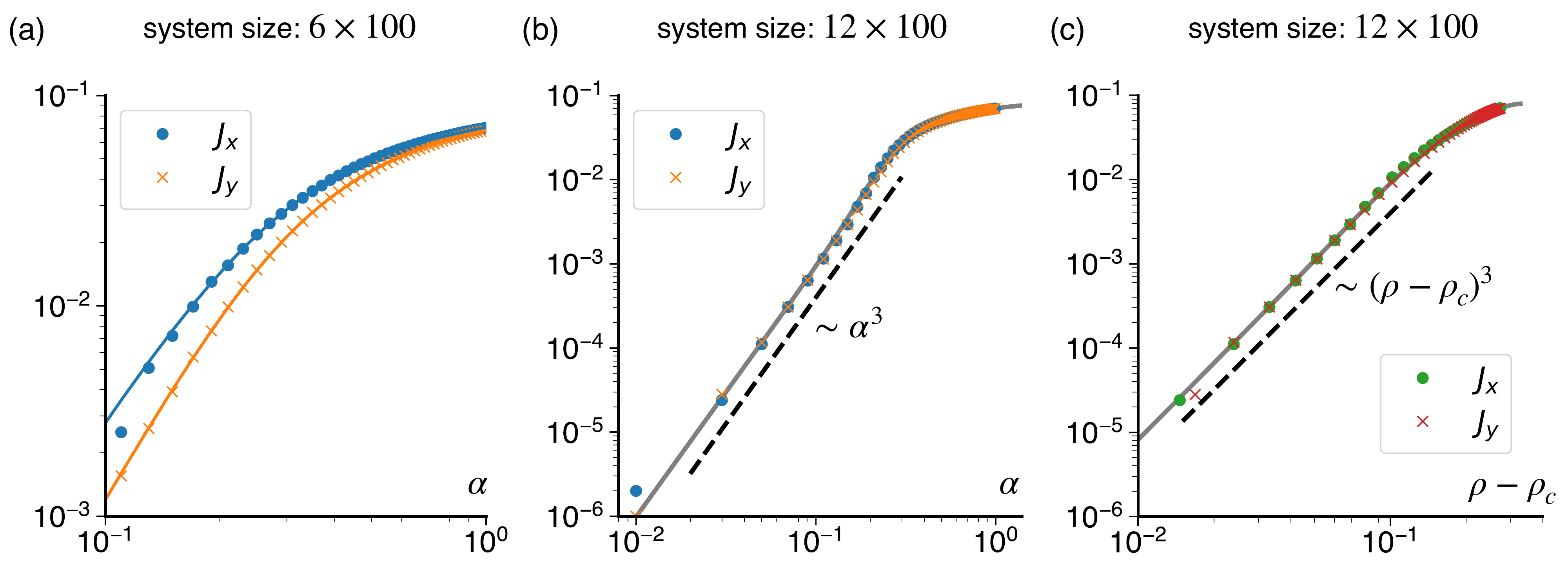}
    \caption{The currents: Log scale plot of   $J_x$ and $J_y$  vs. $\alpha$   for (a) $6\times 100$ system, (b)  $12\times 100$  system.  For smaller systems, currents are asymmetric in  $x$ and $y$ directions.  For larger systems   the asymmetry disappears  and they exhibit  $\alpha^3$  behaviour near $\alpha =0.$.
    (c)  $J_{x,y}$ vs. $\rho-\rho_c$ also  shows  a power-law  near $\rho=\rho_c=\frac12$. Symbols are data points obtained from Monte-Carlo simulations and solid lines are exact MPA solutions. The dashed lines highlight the scaling behaviour.}
    \label{fig:Jxy}
\end{figure*}

In Fig.~\ref{fig:Jxy}(a), we plot $J_{x,y}$ versus $\alpha$, obtained from Monte Carlo simulations of a $6\times100$ system (symbols), together with the analytical results calculated in the $L\to\infty$ limit (lines). The data agree well with the theoretical predictions. Since particle motion is more restricted in the $y$-direction, the corresponding current satisfies $J_y<J_x$ for all values of $\alpha$.
The difference between $J_x$ and $J_y$ decreases as $M$ increases; already for $M=12$ it becomes negligible, as shown in Fig.~\ref{fig:Jxy}(b). There, the $J_{x,y}$ versus $\alpha$ data for a $12\times100$ system, plotted on a log-log scale, collapse onto each other and match the theoretical curve (line).

For small $\alpha$, we observe a power-law behavior $J_{x,y}\sim\alpha^3$. The origin of this scaling is that the birth--death ratio $\alpha$ acts as a fugacity in the grand-canonical ensemble (see Eq.~\eqref{eq:GCE}). A particle can move only when three neighbouring sites are occupied, which introduces a factor $\alpha^3$. Since for small $\alpha$, the density behaves linearly as $\rho\simeq\frac12+\frac{\alpha}{2}$, one expects the scaling $J_{x,y}\sim(\rho-\frac12)^\beta$ with $\beta=3$. A log-scale plot of $J_{x,y}$ as a function of $\rho(\alpha)-\frac12$ confirms this behavior in Fig.~\ref{fig:Jxy}(b).

In the NAEM, the density is controlled externally by the birth--death ratio $\alpha=\frac{p}{q}$. We will see below that when $p=q=0$, the density becomes conserved and acts as an external control parameter, leading to an absorbing phase transition at a critical density $\rho_c=\frac12$. In this case, $\beta=3$ emerges as the order-parameter exponent, with the activity $\rho_a=2(J_x+J_y)$ serving as the order parameter.

\subsection{Conserved dynamics: $p=0=q$}
In the absence of nonconserved dynamics ($p=q=0$), particles can only move to a vacant neighbouring site with rate $r$ if and only if all other neighbouring sites are occupied. Thus,  a particle is called  active  when  three  of its neighbours are occupied and one neighbour is vacant (so that  it can move following the conserved dynamics). A configuration containing at least one active particle is  an active configuration, else (when there  is  no active particles)  it is  absorbing. 

Since the dynamics is  particle conserving and can never increase the number of consecutive zeros, the system evolves toward configurations that are free of $00$ pairs in the $t\to\infty$ limit. The minimum density of such configurations is $\frac12$. For densities $\rho<\frac12$, configurations without consecutive zeros are impossible, and the system inevitably reaches an absorbing state. Thus the density $\rho$ acts as an external control parameter, and the system is expected to undergo an absorbing phase transition (APT) at the critical density $\rho_c=\frac12$.

It is important to note that even in the active phase ($\rho>\frac12$), not all configurations without $00$ pairs are active. For example, at density $\rho=\dfrac{\nu}{\nu+1}$, $\nu\in \mathbb{N}$, one can construct repeating configurations in which no particle has exactly one vacant neighbour; such configurations are therefore absorbing.
Some  example of such absorbing configurations  having  particle density  (a)~$\rho=\dfrac12$  (b)~  $\rho=\dfrac23$ (c)~$\rho=\dfrac56$ are shown below. 
\bea
\boxed{\begin{array}{c}
..0101..\\
..1010..\\
..0101..\\
..1010..\\
..0101..\\
..1010..\\
\end{array}} ~~~~~~~
 \boxed{\begin{array}{c}
..011011..\\
..101101..\\
..110110..\\
..011011..\\
..101101..\\
..110110..\\
\end{array}}~~~~~~~
\boxed{\begin{array}{c}
..011111..\\
..101111..\\
..110111..\\
..111011..\\
..111101..\\
..111110..\\
\end{array}} ~~~~~~~
 \boxed{\begin{array}{c}
..01..011111..\\
..10..101111..\\
..01..110111..\\
..10..111011..\\
..01..111101..\\
..10..111110..
\end{array} }\\
\hspace*{-.4 cm}(a)~ \rho=\frac12 ~~~~~~~~~~ (b)~  \rho=\frac23~~~~~~~~~~~ (c)  ~\rho=\frac56 ~~~~~~~~(d)~ \rho=x\frac12 +(1-x)\frac56 \nonumber
\eea
One can also combine a $\rho=\dfrac12$ configuration with configurations of density $\rho=\dfrac{\nu}{\nu+1}$ to construct absorbing states with any density $\rho>\frac12$. For example, placing a $\rho=\frac12$ configuration next to a $\rho=\frac56$ configuration (as shown in (d)) does not generate any activity; the resulting state remains absorbing and has an overall density satisfying $\frac12<\rho<\frac56$. More generally, in a large $M\times L$ system (with large $M$ and $L$), absorbing configurations with arbitrary density can be constructed by choosing an area fraction $x$ with density $\rho=\frac12$ and a fraction $1-x$ with density $\rho=\frac{\nu}{\nu+1}$. The overall density then becomes $\rho=(1-x)\dfrac{\nu}{\nu+1}+\dfrac12 x$, which can approach unity for large $\nu$. Moreover, absorbing configurations are not limited to densities of the form $\frac{\nu}{\nu+1}$; other rational densities can also be realized.

The key point is that, although absorbing configurations exist for any density $\rho>\frac12$, they are not dynamically accessible. Starting from an active configuration without $00$ pairs, the conserved dynamics cannot lead the system to an absorbing state.
In the absence of $00$s, any allowed particle move can be reversed with the same rate; therefore, as in the NAEM, all active steady-state configurations are equiprobable. The canonical partition function of the system is then 
\be
Q_N =  \sum\limits_{\{n_{ij} \}}  R(\{n_{ij} \}) \delta\left( \sum_{ij}^L n_{ij} -N\right)
\ee
 where $R(\cdot)$ is an indicator function that equals $0$ if the configuration $\{n_{ij}\}$ contains one or more $00$ pairs, and $1$ otherwise. This definition formally overcounts configurations, since some of them are not dynamically accessible. We assume, however, that in the thermodynamic limit such configurations form a measure-zero subset compared to the recurrent configurations that are mutually accessible. In other words, the ergodic sector effectively covers almost all configurations.

The grand partition function of the system is then
\be 
{\cal Z} = \sum_{N=0}^\infty z^N Q_N,
\ee
which has the same form as Eq.~\eqref{eq:GCE} upon replacing $z$ by $\alpha$. Near the critical density $\rho_c=\frac12$, the density behaves as $\rho(z)=\frac{1+z}{2}$, implying $z\sim\rho-\rho_c$. This relation can be used to determine the order-parameter, $\rho_a$, exponent $\beta$ at the absorbing phase transition.

The density of active particles $\rho_a$ equals the probability of finding an occupied site with \underline{exactly} three occupied neighbours; hence the fugacity contribution scales as $z^3$. One therefore expects $\rho_a\sim(\rho-\rho_c)^\beta$ with $\beta=3$.
In Fig.~\ref{fig:Jxy}(a), we  have plotted  $J_x$ and $J_y$ as functions of $\alpha$, whereas in Fig.~\ref{fig:Jxy}(b), $\rho_a$ as a function of $\rho-\rho_c$, with $\rho_c = 1/2$. For large $M$, we unsurprisingly find that $J_x=J_y=J$. Since in the grand-canonical ensemble of the conserved system, $z\equiv \alpha$ and  $\rho_a = 2(J_x+J_y)=4J$, Fig.~\ref{fig:Jxy}(c) confirms that the order-parameter exponent of the absorbing transition is indeed  $\beta=3$.

\subsection{No move dynamics: $r=0$}

Another interesting limit of the dynamics is $r=0$, where only birth--death processes are present and particle hopping is absent. In this case, it is easy to see that the dynamics remains ergodic within the subspace of configurations that contain no consecutive $00$ pairs.
The steady state of the system in this limit is identical to that of the case with $r\neq 0$, since all allowed configurations (i.e., those without $00$ pairs) are still sampled uniformly by the dynamics. Consequently, the steady-state measure and all thermodynamic quantities remain unchanged.
\subsection{Connection to hard-square lattice gas}

The NAEM model discussed here is closely related to the hard-square lattice gas (HSLG) with infinite nearest-neighbour repulsion, also known as the first-nearest-neighbour exclusion model (1-NN)~\cite{Fisher1963, Gaunt1965}. The HSLG is one of the simplest systems exhibiting entropy-driven ordering in two dimensions.
The model is defined on a two-dimensional square lattice where each site can be occupied by at most one particle, subject to the constraint that nearest-neighbour sites cannot be simultaneously occupied. Since all allowed configurations have equal energy, the thermodynamics is governed purely by configurational entropy and the fugacity $z$, which controls the particle density $\eta=\langle N\rangle/L^2$.
As the density increases, random placement of particles increasingly restricts available sites, and the system gains entropy by preferentially occupying one of the two sublattices. Eventually, perfect sublattice order is achieved at the maximum possible density $\eta=\frac12$.

Initial theoretical descriptions of the HSLG were developed using Bethe-lattice and mean-field approaches by Burley~\cite{Burley1961} and later by Runnels~\cite{Runnels1967}, while fundamental insights into lattice-gas phase transitions were provided by Fisher~\cite{Fisher1963}. Transfer-matrix and cluster-variational studies in the 1960s by Temperley and Runnels~\cite{Temperley1962,Runnels1965} established the existence of an order--disorder transition associated with sublattice symmetry breaking; this transition belongs to the Ising universality class. Subsequent progress relied on increasingly accurate analytical and numerical methods, including series expansions and transfer-matrix calculations~\cite{Baxter1980}, renormalization-group analyses~\cite{Racz1980}, and large-scale Monte Carlo simulations~\cite{BinderLandau1980, Fernandes2007}. Mathematical and combinatorial aspects were later summarized in the review by Runnels~\cite{Runnels1972}, while modern density-functional approaches further refined the thermodynamic description~\cite{LafuenteCuesta2003}. More recently, this mechanism Aof entropy-driven ordering has been placed in a broader modern framework~\cite{Han2025}.

Surprisingly, despite its apparent simplicity as a counting problem, the exact values of the critical fugacity $z_c$ and critical density $\eta_c$ remain unknown. The best available estimates still trace back to early work by Baxter et al.~\cite{Baxter1980},
\be 
z_c=3.7962  ~~~ \eta_c= 0.3677.
\ee
In a  later work \cite{Baxter99} Baxter   pointed out that  the  value of $\eta$    at $z=1$ can be   calculated  to   incredible  numerical accuracy,  
\be 
 \eta({z=1})= 0.22657081546271468894199226347129902640080.
 \label{eq:rho0-41decimal}
\ee

In the NAEM, consecutive $0$s cannot be placed on neighbouring sites; thus, the steady state of the model corresponds to the hard-square lattice gas (HSLG) under the mapping $1 \leftrightarrow 0$. The mean density is therefore $\rho = 1-\eta$, and its minimum value $\rho=\frac12$ corresponds to the maximum packing density of the HSLG.
We thus expect that the density of $0$s in the NAEM at $\alpha_c=\frac{1}{z_c}=0.263421$ should match $\eta_c$. From Eq.~\eqref{eq:rho23}, we obtain
\be
1-\rho\!\left(\alpha=\frac{1}{3.7962}\right)=0.36312,
\ee
which is reasonably close to the corresponding estimate for the HSLG.
The values $\alpha_c=0.263421$ and $\rho_c=1-\eta_c=0.6323$ appear to coincide with the crossing point of the $\rho(\alpha)$ curves (see Fig.~\ref{fig:rho_M}(c)) for different $M$. However, we do not yet have a clear understanding of this behavior or how to exploit it to determine $\rho_c$ more accurately.

As mentioned earlier, the matrices obtained using the MPA are exact; however, in evaluating the partition function we have used an approximation. Since $\rho_0$ can be computed exactly for small $M$, we now evaluate it for $z=1$.
To calculate $\rho_0$, we use Eq.~\eqref{eq:projector_method} with $\alpha=1$. At this value, ${\mathbb T}_M$ becomes a binary matrix, and its largest eigenvalue and eigenvectors can be computed accurately. Straightforward numerical diagonalization yields
\be
\rho_0 = 0.231618,\; 0.227577,\; 0.226775,\; 0.226613,\; 0.22658,\; 0.226573,\; 0.226571
\ee
for $M=4,6,8,10,12,14,16$, respectively, in good agreement with Eq.~\eqref{eq:rho0-41decimal}.

\section{Conclusion}
We have developed a two-dimensional matrix-product ansatz (MPA) that extends the scope of exact steady-state constructions beyond one-dimensional stochastic systems. The formalism provides a general algebraic framework for treating interacting nonequilibrium stochastic processes in two dimensions. As a representative example, we introduce the nonconserved assisted exclusion model (NAEM) and obtain the exact steady-state probabilities of configurations. In the NAEM, particles with exactly one neighbouring site vacant can move there with rate $r$, while birth and death processes occur with rates $p$ and $q$, respectively, only at sites whose neighbouring sites are all occupied. The MPA formalism yields exact steady-state weights and enables a systematic computation of observables using projector techniques, allowing us to obtain explicit results for the mean and higher moments of the particle density and currents. In finite geometries, we show that the current exhibits an anisotropy between the $x$ and $y$ directions, which vanishes in the thermodynamic limit.

The model  in the particle-conserving limit, undergoes an absorbing phase transition at the critical density $\rho_c=\frac12$, with the activity serving as the order parameter. From the  steady state results   we find that the transition is characterized by a nontrivial order parameter exponent $\beta=3$. We have also shown that the steady state of the NAEM can be mapped to the hard-square lattice gas with nearest-neighbour exclusion, thereby establishing a direct connection between nonequilibrium dynamics and an equilibrium entropy-driven ordering problem. 

Our results demonstrate that two-dimensional matrix-product ansatz  provide a powerful and systematic route to exact solutions in interacting stochastic systems, and open up possibilities for studying a broader class of nonequilibrium models with constrained dynamics and emergent critical behavior.

\ack
{P.K.M. acknowledges the financial support provided by ANRF, Science and Engineering Research Board (SERB), DST, Government of India, under Grant No. MTR/2023/000644. P.K.M. also acknowledges MPI-PKS for supporting his visit, during which part of this work was carried out. C.G.R. gratefully acknowledges funding from the Max Planck Society. A.B. acknowledges Priyajit Jana and Indranil Mukherjee for helpful discussions. A.B. also gratefully acknowledges the Bangalore School of Statistical Physics (BSSP)-XV, organised by the International Centre for Theoretical Sciences, Bengaluru, and the Raman Research Institute, Bengaluru, where this collaboration was initiated.}
\section*{References}
\bibliographystyle{iopart-num}
\providecommand{\newblock}{}

\end{document}